\documentclass[review]{elsarticle}

\usepackage{lineno,hyperref}
\journal{Journal of \LaTeX\ Templates}

\usepackage[utf8]{inputenc} 
\usepackage[T1]{fontenc}
\usepackage{url}
\usepackage{ifthen}
\usepackage{amssymb}
\usepackage[cmex10]{amsmath} % Use the [cmex10] option to ensure complicance
\usepackage{epsfig}
\usepackage{amsfonts}
\usepackage{pict2e}
\usepackage{color}

\def\qed{\hfill $\blacksquare$}
\newtheorem{Theorem}{Theorem}
\newtheorem{Definition}{Definition}
\newtheorem{Lemma}{Lemma}

\newtheorem{Proposition}{Proposition}

\newtheorem{Example}{Example}

\def\M{{\mathcal M}}

\def\Q{{\mathcal Q}}

\def\X{{\mathcal X}}
\def\Y{{\mathcal Y}}

\begin{document}
%\LARGE
\begin{frontmatter}
\title{Algorithms for $q$-ary Error-Correcting Codes with Limited Magnitude and Feedback}
%\tnotetext[mytitlenote]{Fully documented templates are available in the elsarticle package on \href{http://www.ctan.org/tex-archive/macros/latex/contrib/elsarticle}{CTAN}.}
%% Group authors per affiliation:
\author{Christian Deppe}
\address{Institute for Communications Engineering\\ 
	                  Technical University of Munich\\
                    D-80333 Munich, Germany\\
                    Email: christian.deppe@tum.de}
										
\author{Vladimir Lebedev}
\address{Kharkevich Institute for Information Transmission Problems\\
	                  Russian Academy of Sciences\\
										127051 Moscow, Russia\\
										Email: lebedev37@mail.ru}
																				
%\fntext[myfootnote]{Since 1880.}

%% or include affiliations in footnotes:

\begin{abstract}
 Berlekamp and Zigangirov completely determined the capacity error function for binary error correcting codes with noiseless
feedback. 
It is still an unsolved problem if the upper bound for the capacity error function
in the non-binary case of Ahlswede, Lebedev, and Deppe is sharp. 
%We present an alternative algorithm
%with partial feedback that reaches the capacity error function for large error fractions. 
We consider wraparound channels with limited magnitude and noiseless feedback.
We completely determine the capacity error function for all $q$-ary wraparound channels with a magnitude of level $r$. All of our algorithms
use partial noiseless feedback. Furthermore, a special case of the problem is equivalent to Shannon's zero-error problem.
\end{abstract}
\end{frontmatter}
%linenumbers
\section{Introduction}\label{introduction}
Error-correcting (block) codes are important in many practical applications. 
For these applications, asymmetric channels are common cases both in fiber and free space. 
Furthermore, in memories like LSI/VLSI, ROM, and RAM, asymmetric errors happen. These are the faults that affect address decoders, word lines, power supplies, and stuck-faults in a serial bus.
An asymptotic information
theoretic analysis of these codes is difficult. The capacity error correcting function is
still unknown, even in the binary case. In this case, one assumes that a channel can change
no more than $t$ symbols and that the number of errors is proportional to the block length.
It is much easier to consider the probabilistic model of a channel where we assume that a
symbol is changed to another symbol with a certain probability. This model was introduced by
Shannon in \cite{S48}. He gave a formula for the channel capacity of the binary case (binary symmetric channel) 
as well as for a more general case with a $q$-ary alphabet (discrete memoryless channel). In 
these cases the channels are discrete and memoryless. Shannon considered 
this model in \cite{S56}, which also has noiseless feedback, and he showed that the capacity is the same.
Ahlswede showed in \cite{A73} that the direct proof becomes easier in this case. Instead of an existence proof, one
gets a constructive proof. In the case of binary error correcting codes, Berlekamp considered error correcting codes with noiseless feedback in \cite{B68}. 
He assumed that the errors are proportional to the block-length.
It turned out that the capacity error correcting function is different from
the one without noiseless feedback. The capacity error function is larger for all error fractions $\tau$ between 0 and $\frac 13$. 
Furthermore, together with a result from Zigangirov in \cite{Z76}, the error fraction is completely known. 
In \cite{D05},
the $q$-ary case is considered. In this case, for $0\leq \tau\leq \frac 1q$, the capacity error function
is not completely known. The authors showed that the rubber strategy reached the upper bound for an error fraction larger than $\frac 1q$. 
%We give in Section~3 a new algorithm
%that also assumes the capacity error correcting function for a error fraction larger $\frac 1q$. Our algorithm uses partial feedback.
%This is the first optimal algorithm with partial feedback when the error is proportional to the block length. Algorithms with partial feedback
%have been analyzed (see \cite{cicalese}) only for a fixed number of errors.
From a practical view it is not always realistic that every symbol be changed to every other symbol. 
Therefore it makes sense to restrict this by introducing a directed bipartite graph in which
the vertices in each part are labeled with the coding alphabet. There is a directed line between symbol $a$ and
$b$ if the channel can change symbol $a$ to symbol $b$. 

For a $q$-ary alphabet, we consider the case that $a$ can be changed to $a\oplus 1$, $a\oplus 2$, $\dots$, $a\oplus r$, where $\oplus$ stands for the addition modulo $q$.
These channels are called wraparound channels with magnitude $r$
(\cite{BET18}). If $r=q-1$,
the graph is a complete bipartite graph, and we are in the $q$-ary case considered in \cite{B68}.
If $r<q-1$, the channel is called a channel with limited magnitude.
We assume for all channels considered that we have noiseless feedback. Our algorithms only partially use this feedback.

This paper is organized as follows.
In Section~2 we give the definitions and known results for the classical $q$-ary case. 
In Section~3 we introduce wraparound channels with magnitude 1 and therefore consider a 2-regular channel
graph. We show that we can calculate the capacity error correcting function 
for all $q>3$ and give a coding algorithm that reaches this function. We generalize this result in Section~4 for wraparound channels with magnitude $r$. In this case an error corresponds to an addition to the symbol sent of a quantity $s$ in $[0,r]$ modulo $q$. Furthermore,
we give a general theorem, which extends the class of graphs for which we can apply our algorithm. Therefore, we give the error correcting capacity and a coding algorithm that reaches this function.     
In Section~5 we give a conclusion and show that our result is related to Shannon's zero-error capacity if $\tau=1$. 
\section{$Q$-ary codes with feedback}\label{codes}
Consider communication over a $q$-ary channel
with input alphabet $\X=\Q=\{0,1,\ldots ,q-1\}$ and output alphabet $\Y=\Q$, where a word of length $n$ sent
by the encoder is changed by the channel in at most $t$ letters.
In our model, a sender wants to transmit a message $m\in\M=\{1,2,\dots, M\}$ over a $q$-ary channel with noiseless feedback. $\M$ is the message set and $\Q$ is the coding alphabet. 
Suppose now that having sent $(x_1,\ldots ,x_{j-1})=x^{j-1}\in\X^{j-1}$, the encoder
knows the received letters $(y_1,\ldots ,y_{j-1})\in\Y^{j-1}$ before it sends the
next letter $x_j$ ($j=1,2,\ldots ,n$). In this case we are dealing with a noiseless feedback channel.
The code words are elements of $\Q^n$. An encoding function (algorithm) is defined by
\[
c(m,y^n)=((c_1(m),c_2(m,y_1),\dots,c_n(m,y^{n-1}),
\]
where $c_i:\M\times\Q^{i-1}\to \Q$ is a function for the $i$th letter, which depends on the message and the $(i-1)$
letters which have been sent before. 
The receiver has a decoding function $d:\Q^n\to \M$, which maps the received
sequence to a message.
The coding scheme is as follows:
\begin{figure}[h]
	\centering
	\setlength{\unitlength}{0.8 cm}
	\begin{picture}(12,6)
	\put(1,1){\framebox(2.5,1){Sender}}
	\put(5,1){\framebox(2.5,1){Channel}}
	\put(9,1){\framebox(2.5,1){Receiver}}
	\put(3.5,1.5){\vector(1,0){1.5}}
	\put(7.5,1.5){\vector(1,0){1.5}}
	\put(8.25,1.5){\line(0,1){3.5}}
	\put(8.25,5){\vector(-1,0){4}}
	\put(2.25,5){\line(1,0){4}}
	\put(2.25,2){\line(0,1){3}}
	\put(4.5,4.5){Feedback}
	\put(5.5,2.5){\framebox(1.5,1){Noise}}
	\put(6.2,2.5){\vector(0,-1){0.5}}
	%\put(6,1.25){\line(1,1){0.5}}
	%\put(6.5,1.75){\vector(0,-1){1.5}}
	\end{picture}
	\caption{Channel with feedback}
	\label{fig:channel_feedback}
\end{figure}
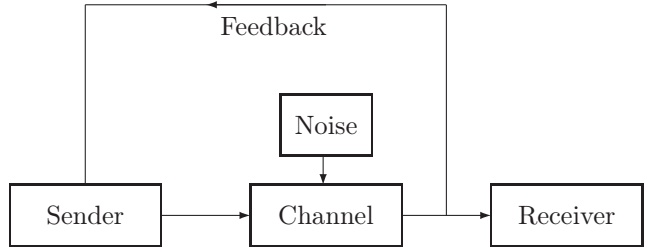
We assume that the noise can change fewer than $t$ symbols. Furthermore,
we assume that $\tau=\frac tn$. This means the number of errors is proportional to the block length.
\begin{Definition} Let $M$ be the number of messages, $n$ be the block-length,
	and $t$ be the maximal number of errors.
We speak about a successful $(n,M,t)_f$ coding algorithm with noiseless
feedback if there exists an encoding function of the sender and a decoding function of the receiver, such 
that the receiver decodes the message of the sender correctly if fewer than $t$
errors happen.
\end{Definition}
In our paper, all described coding algorithms considered are successful.
For $q=2$, this model was considered by Berlekamp \cite{B68}, who gives successful $(n,M,t)_f$ coding algorithms with noiseless feedback. 
\begin{Definition}
	Let $M$, $n$, $t=n\tau$, be defined as above. The rate of a successful coding
	algorithm is defined as 
	\[
	R=\frac {\log_q M}{n}.
	\]
The supremum of the rates of a successful algorithm achievable for
$\tau$ and all large $n$ is called {\bf the capacity error function 
(or curve)} with feedback, and is denoted by $C^f_q(\tau )$.  
\end{Definition}
In \cite{B68} and \cite{Z76}, $C^f_2(\tau)$ was determined. It was shown that
\[ 
C^f_2(\tau)= \left\{\begin{array}{ll} 1-h(\tau) &\text{if } 0\leq \tau\leq ({3+\sqrt 5})^{-1}, \\
(1-3\tau) \log \left(\frac{1+\sqrt 5}{2}\right) &\text{if } ({3+\sqrt 5})^{-1}\leq \tau\leq \frac 13,
\end{array}\right.
\]
where $h(\tau)=-\tau\log_2\tau-(1-\tau)\log_2(1-\tau)$ denotes the binary entropy.
Furthermore, we have $C^f_2(\tau)=0$, if $\tau>\frac 13$.
\bigskip
In the case where $q>2$, we have the following bound (see \cite{D05}):
\begin{eqnarray*}%\label{eq:1}
C^f_q(\tau )\leq
\begin{cases}
1- h(\tau ) \log_q 2 - \tau\log_q(q-1), & 
\text{if\ } 0\leq \tau \leq \frac{1}q,\\
(1-2\tau)\log_q(q-1) & \text{if\ } \frac{1}q < \tau \leq \frac 12,\\
0 & \text{if\ } \tau > \frac 12.
\end{cases}
\end{eqnarray*}
In \cite{D05}, the so-called 1-rubber method is used to show that for $\frac 1q\leq \tau\leq \frac 12$,
\[
C^f_q(\tau )=(1-2\tau)\log_q(q-1).
\]
In the 1-rubber method,
the receiver just regards the ``0'' as
a protocol symbol - it erases it by a rubber, which in addition erases the previous symbol.
The symbols in the set $\{1,\dots,q-1\}$ are used as information symbols, which means 
the sender maps each message to one sequence of information symbols. The decoder knows this 
mapping and can get the message by recovering the sequence of information symbols.
In the 1-rubber method the encoder doesn't use ``0'' as an information symbol and $M=(q-1)^{n-2t}$. 
\begin{Example}[1-rubber method of \cite{D05}]
Let $n=5$, $t=2$, $q=3$, and $b:\M\to\{1,2\}$ be the map which maps the messages to the information symbols: 
\[
b(1)=1,\ \ \ b(2)= 2.
\]
The sender wants to send message 1.\\
\begin{center}
\begin{tabular}{rccc}\rm
sent: & {\bf 1}\rm 0111 & {\bf 1}\rm 0{\bf 1}01 & {\bf 1}\rm{\bf 0}001\\
received: & \rm \underline{{\bf 2}0}111 & \rm \underline{{\bf 2}0}\underline{{\bf 2}0}1 & \rm \underline{{\bf 2}\underline{{\bf 2}0}0}1\\
\end{tabular}
\end{center}
\end{Example}
The capacity error function of the 1-rubber method can 
be calculated: 
\[ R(\tau)=(1-2\tau)\log_q(q-1).\]
It is a tangent to the capacity error function $C^f_q$ if $q>3$. It goes through $(1/2,0)$ and touches $C^f_q$ at $(1/q,\frac{q-2}q \log_q (q-1))$.
The 1-rubber method gives a positive capacity only for $q>2$, because for $q=2$ we can transmit only one message.
It was also proved that instead of a single zero, one can use
$a$ consecutive
zeros for erasing. This asymptotically gives tangent lines to $C^f_q$ 
going through the points $(1/(a+1),0)$. This method is called the $a$-rubber method. This method also works for $q=2$ if $r\geq 2$.
The 2-rubber method achieves the same rate as Berlekamps optimal algorithm in the binary case. More details of the rubber method can be found in \cite{D05}.
\section{An algorithm for 2-regular channel graphs}\label{one}
\begin{figure}[h]
\setlength{\unitlength}{1 cm}
\begin{center}
\begin{picture}(8,3.5)
\put(1.5,3){0}
\put(1.5,2.2){1}
\put(1.5,1.3){2}
\put(1.5,0.6){$\vdots$}
\put(1,0.1){$q-1$}
\put(6,3){0}
\put(6,2.2){1}
\put(6,1.3){2}
\put(6,0.6){$\vdots$}
\put(6,0.1){$q-1$}
\put(2,3.15){\line(1,0){3.5}}
\put(2,2.26){\line(1,0){3.5}}
\put(2,1.36){\line(1,0){3.5}}
\put(2,0.25){\line(1,0){3.5}}
\put(2,3.15){\line(4,-1){3.5}}
\put(2,2.26){\line(4,-1){3.5}}
\put(2,0.25){\line(6,5){3.5}}
\end{picture}
\end{center}
\caption{Channel graph of a $q$-ary channel with limited magnitude}\label{magnitudefig}
\end{figure}
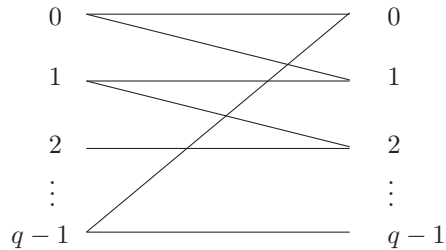

The special $q$-ary channel we consider here can be presented by a bipartite graph of Figure~\ref{magnitudefig}.
\bigskip
The lines show which symbol on the right side could be received if a symbol on the left side is transmitted.
%Therefore an error corresponds by adding 1 to a send symbol (mod $q$).
We denote by $\Gamma_1(q)$ the set of all channel graphs in which there is only one error possible per symbol.
Another way to represent the channel is a $q\times q$ matrix $(a_{ij})_{0\leq i,j \leq q-1}$, like in Figure~\ref{matrix}. This is the adjacency matrix of the graph.
The value $a_{ij}$ in row $i$ column $j$ is equal to 1 if an error can occur from $i$ to $j$ and $a_{ii}=1$, since we consider channels 
where we assume that the symbol is transmitted correctly, if no error occurred. 
Otherwise, the entry is 0.
\begin{figure}[h]\label{matrix}
\[
\left( \begin{array}{rrrr}1 & 1 & 0 & 0 \\0 & 1 & 1 & 0 \\0 & 0 & 1 & 1 \\1 & 0 & 0 & 1 \\\end{array}\right) 
\]
\caption{$Q$-ary channel with limited magnitude for $q=4$ in matrix representation}
\end{figure}
All graphs (channels) in $\Gamma_1(q)$ have in their matrix representation only 1s in the diagonal and another 1 in each column. All other entries are 0.
If all values of a channel matrix are 1, or the channel graph is a complete bipartite graph, we have the case of a $q$-ary error correcting code with feedback without the restrictions of Section~\ref{codes}.

We now consider the set $\tilde{\Gamma}_1(q)\subset\Gamma_1(q)$. Every vertex
of every graph in $\tilde{\Gamma}_1(q)$ has a degree of 2. This means if we know the position of an error we can correct it.
To give the capacity error function for the channels in $\tilde{\Gamma}_1(q)$ we need the following lemmas.
\begin{Lemma}\label{D}
Let $\Delta\in \tilde{\Gamma}_1(q)$. If there exists a successful algorithm for $\Delta$ with rate $R_1(\tau)>0$ for any $\tau$, then 
there is an successful algorithm $D_1(\Delta)$ with rate $1-\log_q 2$ for any $\tau$.
\end{Lemma}
{\bf Proof:} According to the assumptions of the lemma, an algorithm with a positive rate exists. If this algorithm has the rate $R_1(\tau) = R_1= 1-\log_q 2$, we are done. So we assume that it has a rate $R_1<1-\log_q 2$. We now construct a new successful algorithm
with rate $R_2$ in the following way.
In the interval $[1,a]$, which means for the symbols $x_1,\dots,x_a$, we send information symbols. The sender knows the received sequence because of the
noiseless feedback. Therefore, in the interval $[a+1,n]$, he has to send the positions where
an error occurred. Then we can correct the information symbols, because in our scheme 
each symbol can only be mapped to one other symbol. To transmit this information we use the algorithm with rate $R_1$ for any number of errors. 
Then if
$$ 2^a = q^{(n-a)R_1},$$
we have an algorithm with rate $R_2$ for any numbers of errors $R_2$ such that
$$R_2= a/n= \frac {R_1}{\log_q2+R_1}.$$
The rate $R_2>R_1$ if $R_1>0$, otherwise $R_2=0$.
Now we can use this algorithm with rate $R_2$ as an algorithm for any number of errors and use the same method as before to 
get an algorithm with rate $R_3$.

Therefore in general, using this idea iteratively, we get for $i\geq 2$,
\[
R_i= \frac {R_{i-1}}{\log_q2+R_{i-1}}.
\]
For a positive constant $c$, the function $f(x)= \frac {x}{x+c}$ is increasing and concave. For $f(x)=x$ we have $x+c=1$; therefore the iterations give us an algorithm with rate 
\[
1-\log_q 2
\] 
for any number of errors.\qed
\begin{Lemma}\label{L}
Let $\Delta\in\tilde{\Gamma}_1(q)$ and let $A$ be a successful algorithm for $\Delta$ and with rate $R_1(\tau)>0$ for any $\tau$, then
we give a successful algorithm $L_1(\Delta)$ with rate $1-h(\tau)\log_q 2$ for $\Delta$ and $0\leq\tau\leq \frac 12$.
\end{Lemma}
{\bf Proof:} Let $R_1(\tau)=R_1=\log_q \gamma$ with $\gamma>1$. We will give an encoding and a decoding algorithm such that we can transmit $\frac {q^n}{{n\choose t}}$ messages; this
algorithm has the rate $1-h(\tau)\log_q 2$.
Let $k_1=n-\log_q{n \choose t}$. 
In the interval \[
S_{k_1}=[1,k_1]=(x_1,x_2,\dots, x_{k_1}),
\] 
we transmit information symbols. We have to show that the receiver can decode this sequence.
Thus we have
\[ n_1=\log_q{n\choose t}\]
symbols to inform the receiver about the $t_1$ errors appearing in $S_{k_1}$. 
If he gets this information, he is able to decode the message correctly.
If ${k_1\choose t_1}\leq \left(\gamma\right)^{n_1}$ we 
can use algorithm $A$, otherwise let $k_2=\log_q{k_1 \choose t_1}$.
In the interval 
\[
S_{k_2}=[k_1+1,k_1+k_2],
\] 
we send the information about the $t_1$ errors in $S_{k_1}$. 
We have 
\[
n_2=n_1-k_2
\]
symbols for the next transmission and so on.

We have
\[
k_{i+1}=\log_q{k_{i} \choose t_i},\ n_{i+1}=n_i-k_{i+1},
\]
and stop this procedure if ${k_i\choose t_i}\leq \left(\gamma\right)^{n_i}.$  
Then we just send $ \log_q\left(\gamma\right)^{n_i}$ symbols, using Algorithm $A$.\\

Therefore, before repeating the procedure we check the condition  
$${k_i\choose t_i}> \left(\gamma\right)^{n_i}.$$
If it is not true for some $i$, then we use Algorithm $A$ and send  $\left(\gamma\right)^{n_i}$ symbols.
\bigskip 
If it is true, then
\[
\frac {n_{i+1}}{n_i}= \frac {n_i-k_{i+1}}{n_i}= 1-\frac {\log_q{ k_i \choose t_i}}{n_i}\leq 1-\log_q\gamma.
\]
So we have
\[
\frac  {n_{i+1}}{n_i}\leq 1-\log_q \gamma.
\]
Thus, the number of steps in the algorithm is not more than $O(\log n)$.
Now we show that for any $i$ we have $k_{i+1}\leq n_i$.\\
For $i=1$ it is true. Let 
\[
n_i=n_{i-1}-k_i=n_{i-2}-k_{i-1}-k_i=\dots=n_1-k_2-k_3-\dots
\]
So we have to prove  $k_2+k_3+\dots+k_{i+1}\leq n_1$. It holds
\[k_2+k_3+\dots+k_{i+1}=\log_q\left( {k_1\choose t_1}{k_2\choose t_2}\cdots {k_i \choose t_i}\right)\leq \log_q{k_1+k_2+\cdots +k_i \choose t_1+t_2+\cdots +t_i}.\]

We have
$\log_q{k_1+k_2+\cdots +k_i \choose t_1+t_2+\cdots +t_i}\leq \log_q {n\choose t}$
for $t<n/2$.
In order for the receiver to be able to decode the message sent, we must additionally send the information about 
$t_1, t_2, \dots$ in the last interval.
The idea is that in the last interval
we use the algorithm $A$, which can correct any number of errors
to additionally transmit information about $t_1, t_2, \dots$. 
For any $t_i$ we need $\log_q(t)$ positions, and the number of $t_i$s is at most $\log_q n$.
Therefore the decoder first decodes the last interval and then knows the length of all intervals and
can decode the message. 
Thus the algorithm $L_1(\Delta)$ is successful. The rate of the algorithm is $1- h(\tau )\log_q 2$.
\qed

The previous lemmas can be used to prove the following.
\begin{Theorem}\label{TA}
Let $\Delta \in \tilde{\Gamma}_1(q)$ and $q>3$, then $\Delta$ has the following capacity error function
\begin{eqnarray*}
C_{\Delta, f}^q(\tau )=
\begin{cases}
1- h(\tau )\log_q 2 & 
\text{if } 0\leq \tau \leq \frac{1}2\\
1-\log_q 2 & \text{if } \tau > \frac 12.
\end{cases}
\end{eqnarray*}
\end{Theorem}
{\bf Proof:} 
We will obtain the upper bound by counting the number of possible output
sequences.
Consider the encoding function
\begin{equation}
\label{eq_8}
c(m,y^n):=(c_1(m),c_2(m,y_1),\dots,c_n(m,y^{n-1}),
\end{equation}
where $c_i:\M\times\Q^{i-1}\to \Q$ is a function for the $i$th letter, which depends on the message and the $(i-1)$
letters which have been sent before. 
The output $y^n = (y_1, \dots, y_n)$ with
\begin{equation}
\label{eq_9}
y_1 = c_1(m) \oplus e_1 \mbox{ and } y_t = c_t(m,(y_1, y_2, \dots y_{t-1}) \oplus e_t \mbox{ for $t = 2,3, \ldots, n$}
\end{equation}
is uniquely determined by the encoding functions $c_i$, the message, and the binary 
error pattern $e^n = (e_1, e_2, \ldots, e_n)$ occurring in the transmission,
and can therefore be regarded as their function $\Phi (c(m,y^n), e^n)$. For a family
of encoding functions and a set ${\cal E}$ of error patterns, we write
\begin{equation}
\label{eq_10}
\Phi (c_{{\cal M}}, {\cal E}) = \{y^n : \mbox{$\exists$ $m \in {\cal M}$
 and $e^n \in {\cal E}$ such that } y^n = \Phi (c(m,y^n), e^n) \}.
\end{equation}
For a unique decoding, for each error pattern we need to get another output sequence, and therefore
\[
M\leq \frac  {q^n}{\sum_{j=0}^t{n\choose j}}.
\]
Let
\[
\tilde{M}(n,t) = \begin{cases} \frac {q^n}{{n\choose t}}  & {\rm if}\  0\leq t \leq \frac n2 \\
\left({\frac q2}\right)^n & {\rm if}\  t >\frac n2. \end{cases}
\]
It is known that 
\[
\lim_{n\to\infty}\tilde{M}(n,t)=\lim_{n\to \infty} \frac  {q^n}{\sum_{j=0}^t{n\choose j}}.
\]
Let $n\tau=t$ and therefore
\begin{eqnarray}\label{upper}
C_{\Delta, f}^q(\tau )\leq
\begin{cases}
1- h(\tau )\log_q 2 & 
\text{if } 0\leq \tau \leq \frac{1}2\\
1-\log_q 2 & \text{if } \tau > \frac 12.
\end{cases}
\end{eqnarray}
Now we give a successful algorithm that attains this bound.
We start with the case $\tau>\frac 12$. 
If $q$ is even, we can transmit $\left(\frac q2\right)^n$ messages for any number of errors,
because we can find a set $S$ of $\left(\frac q2\right)$ symbols so that no error of one of the symbols in $S$ will result in a symbol of $S$. 
This idea was first used by Shannon in \cite{S56}.
It is obvious that this algorithm has the rate $1-\log_q 2$. For $q>3$ this algorithm can also be used for odd
$q$ with rate $1-\log_{q} {\frac {q-1}2}$. Applying Lemma~\ref{D}, we know that there is also for odd $q>3$ an algorithm $D_1(\Delta)$
with rate $\log_q{\frac q2}=1-\log_q 2$.
\begin{figure}[h]
\setlength{\unitlength}{1 cm}
\begin{center}
\begin{picture}(7,3)
\put(1.5,2.5){\textcolor{green}0}
\put(1.5,1.7){1}
\put(1.5,0.8){\textcolor{red}2}
\put(1.5,0){3}
\put(6,2.5){\textcolor{green}0}
\put(6,1.7){\textcolor{green}1}
\put(6,0.8){\textcolor{red}2}
\put(6,0){\textcolor{red}3}
\put(2,2.65){\line(1,0){3.5}}
\put(2,1.76){\line(1,0){3.5}}
\put(2,0.86){\line(1,0){3.5}}
\put(2,-0.04){\line(1,0){3.5}}
\put(2,2.65){\line(4,-1){3.5}}
\put(2,1.76){\line(4,-1){3.5}}
\put(2,0.86){\line(4,-1){3.5}}
\put(2,-0.04){\line(6,4.6){3.5}}
\end{picture}
\end{center}
\caption{Example of Shannon's algorithm, $q=4$, and  $\tau>\frac 12$.}
\end{figure}
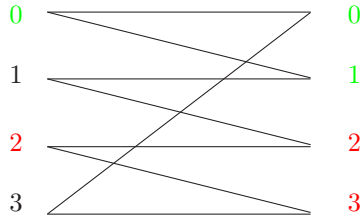
For the case $0\leq \tau\leq \frac 12$ and $q>3$ we can use algorithm $L_1(\Delta)$ of Lemma~\ref{L}, because the assumption
are fulfilled by the previous considerations. We get the rate $1- h(\tau )\log_q 2$.
\qed
%The situation is different, if we have $q=3$
%\begin{Theorem}
%The channel $\Gamma$ has for $q=3$ the following capacity error function.
%\begin{eqnarray*}
%C_f^q(\tau )=
%\begin{cases}
%1- h_q(\tau ) & 
%\text{if } 0\leq \tau \leq \frac{1}2\\
%0 & \text{if } \tau > \frac 12,
%\end{cases}
%\end{eqnarray*}
%\end{Theorem}
%{\bf Proof} The upper bound and the lower bound for the case $\tau\leq \frac 12$ works like in the previous Theorem.
%In the case $\tau>\frac 12$ we cannot use the algorithm $B$ because there is no algorithm with a positive rate.
%This is easy to see, because if we take two arbitrarily sequences and send them over the channel. The channel can
%always modify the sequences in such a way that the receiver cannot distinguish them. \qed%
\section{Generalized algorithms}\label{r-asym}
%???????????, ??? ?????? ??????????? ?????? ?? ?????, ??? ?? $r-1$ (?? ?????? $q$).
%??????? ???????? ????? ????????? ???
In this section we generalize the previous algorithm. Now we consider the channel set $\Gamma_r(q)$ of $(r+1)$-regular graphs. 
For this channel we have $r$ possible
errors for each symbol. In the matrix representation we have $r+1$ 1s per column (including the 1 in the diagonal for correct transmission).  
We give some generalizations of Lemma~\ref{D} and Lemma~\ref{L}.
\begin{Lemma}\label{D_r}
Let $\Delta\in \Gamma_r(q)$, if there exists a successful algorithm for $\Delta$ with a positive rate for any $\tau$, then
there is an algorithm $D_r(\Delta)$ with rate $1-\log_q (r+1)$ for any $\tau$.
\end{Lemma}
{\bf Proof:}
Let us assume that the existing algorithm has a rate $0<R_1'<1-\log_q (r+1)$. We now construct a new algorithm
with rate $R_2'$ in the following way. 
In $[1,a]$ we send information symbols.
Then we use the algorithm with rate $R_1'$ for any number of errors.   
We have to inform the receiver about the error position and the value of the error ($r$ possibilities).
Therefore we choose $a$ such that
$$ \sum_{i=0}^{t} {n\choose i} r^i =(r+1)^a = q^{(n-a)R_1'}.$$
This gives a working algorithm, because we can send all information
about errors to the receiver in the second interval. Therefore, we have an algorithm with rate $R_2'$ for any numbers of errors.

Now we have.
$$ R'_2=\frac an = \frac {R_1'}{R_1'+\log_q (r+1)}.$$
%??? ????? ????????????? ???????? $a$ $b$ ??????? $f(x)= \frac {ax}{b+ax}$ ?? %??????? $[0,1]$ ??????????, ??????? ????? ? ???????????? ? ?????? $g(x)=x$ ? ?????
%$((a-b)/a , (a-b)/a)$.
We can use this algorithm with rate $R'_2$ as an algorithm for any number of errors and use the same method as before to 
get an algorithm with rate $R'_3$.

Therefore, in general, using this idea iteratively, we get for $i\geq 2$,
\[
R_i= \frac {R_{i-1}'}{\log_q (r+1)+R_{i-1}'}.
\]
We use for a positive constant $c$, the function $f(x)= \frac {x}{x+c}$. This function is increasing and concave. For $f(x)=x$ we have $x+c=1$. The iterations give us an algorithm with rate 
\[
1-\log_q (r+1)
\] 
for any number of errors.
\qed
\begin{Lemma}\label{L_r}
Let $\Delta\in\Gamma_r(q)$ and let $A_r$ be an algorithm for $\Delta$ with rate $R_1'(\tau)>0$ for any $\tau$, then
there exists a successful algorithm $L_r(\Delta)$ with rate $1-h(\tau)\log_q (r+1)$ for $\Delta$ and $0\leq\tau\leq \frac r{r+1}$.
\end{Lemma}
{\bf Proof:} 
Let $R_1'(\tau)=R_1'=\log_q \gamma_r$ with $\gamma_r>1$. 
Let $k_1=n-\log_q\left({n \choose t}r^t\right)$.
In the interval $S_{k_1}=[1,k_1]$ we transmit information symbols.
Thus we have
\[ n_1=\log_q \left({n\choose t}r^t\right)\]
symbols for the next transmission. In this transmission 
we want to inform the receiver about the $t_1$ errors appearing in $S_{k_1}$, 
the error positions in the transmission 
of the information symbols and their values. If he gets this information, he is able to decode the message correctly.
We can use the algorithm $A_1$ for this, if ${n\choose t_1}r^{t_1}\leq \left(\gamma_r^{n_1} \right)$.
Otherwise, let $k_2=\log_q\left({k_1 \choose t_1}r^{t_1}\right)$.
In the interval $S_{k_2}=[k_1+1,k_1+k_2]$, we send the information about the $t_2$ errors in $S_{k_1}$.
We have 
\[
n_2=n_1-k_2
\]
symbols for the next transmission.
We have
\[k_{i+1}=\log_q\left({k_{i} \choose t_i}r^{t_i}\right),\ n_{i+1}=n_i-k_{i+1}.
\]
We stop this procedure if ${k_i\choose t_i}r^{t_i}\leq \left(\gamma_r\right)^{n_i}.$
Then we just send $ \log_q\left(\gamma_r\right)^{n_i}$ symbols, using algorithm $A_r$.\\
Therefore, with every step we have to check the condition  
$${k_i\choose t_i}r^{t_i}> \left(\gamma\right)^{n_i}.$$
If it is not true for some $i$, then we use algorithm $A_r$ 
and send  $\gamma_r^{n_i}$ symbols.

Now we show that for any $i$: $k_{i+1}\leq n_i$. For $i=1$ it is true, because ${k_1\choose t_1}r^{t_1}\leq {n\choose t}r^t$ for $t\leq n \frac {r}{r+1}$.
\[
n_i=n_{i-1}-k_i=n_{i-2}-k_{i-1}-k_i=\dots=n_1-k_2-k_3-\dots
\]
So we have to prove  $k_2+k_3+\dots+k_{i+1}\leq n_1$
\begin{eqnarray}
k_2+k_3+\dots+k_{i+1}&=&\log_q\left( {k_1\choose t_1}{k_2\choose t_2}\cdots {k_i \choose t_i} r^{t_1}\dots r^{t_i} \right)\\
&\leq& \log_q\left({k_1+k_2+\cdots +k_i \choose t_1+t_2+\cdots +t_i} r^{t_1+\cdots +t_i}\right).
\end{eqnarray}
We have
\[
\log_q\left({k_1+k_2+\cdots +k_i \choose t_1+t_2+\cdots +t_i} r^{t_1+\cdots +t_i}\right)\leq \log_q \left( {n\choose t} r^t\right).
\]
Thus the encoding works if $\tau \leq \frac{r}{r+1}$ and its rate is 
\[
1- h(\tau)\log_q 2-\tau \log_q r.
\]
It remains to be shown that there is also a valid decoding algorithm. The idea is that in the last interval 
we use the algorithm $A_r$, which can correct any number of errors to transmit additional information about $t_1, t_2, \dots$. 
For any $t_i$ we need $\log_q(t)$ positions and we have at most $\log_q n$ $t_i$s.
Therefore the decoder first decodes the last interval and then knows the length of all intervals and
can decode the message. 
\qed

Now we consider a special subset of $\Gamma_r(q)$. We assume that in the matrix representation of 
the channel in each row there are $r+1$ consecutive 1s including the 1 on the diagonal. We
call this set of channels $\Lambda_r(q)$. 
\begin{Theorem}\label{two}
Let $1\leq r \leq \frac q2-1$ and $\Delta\in \Lambda_r(q)$, then $\Delta$ has the following capacity error function.
\begin{eqnarray*}
C_{\Delta,f}^q(\tau )=
\begin{cases}
1- h(\tau )\log_q 2-\tau \log_q r & 
\text{if } 0\leq \tau \leq \frac{r}{r+1},\\
1-\log_q(r+1) & \text{if } \tau > \frac {r}{r+1}.
\end{cases}
\end{eqnarray*}
\end{Theorem}
{\bf Proof:}
We get the upper bound by counting the number of output sequences for each message. 
The difference to the proof of Theorem~\ref{TA} is that we
have $r$ possibilities for each error.
We get the following:
\[
M\leq \frac  {q^n}{\sum_{j=0}^t{n\choose j} r^j}.
\]
Let
\[
\tilde{M}_r(n,t) = \begin{cases} \frac {q^n}{{n\choose t}r^t}  & {\rm if}\  0\leq t \leq \frac r{r+1} n,\\
\left({\frac q{r+1}}\right)^n & {\rm if}\  t >\frac r{r+1} n. \end{cases}
\]
It is known that
\[
\lim_{n\to\infty}\tilde{M}(n,t)=\lim_{n\to \infty} \frac  {q^n}{\sum_{j=0}^t{n\choose j}r^j}.
\]
From this it follows directly that for $n\to\infty$ and $\tau=\frac tn$, it holds,
\begin{eqnarray}\label{generalupper_r}
C_{\Delta,f}^q(\tau )\leq
\begin{cases}
1- h(\tau )\log_q 2-\tau \log_q r & 
\text{if } 0\leq \tau \leq \frac{r}{r+1}\\
1-\log_q(r+1) & \text{if } \tau > \frac {r}{r+1}.
\end{cases}
\end{eqnarray}
The second line follows from the fact that
\begin{eqnarray*}
& &1-h_q(\frac {r}{r+1})-\frac {r}{r+1} \log_q r\\
&= & 1+\frac r{r+1}\log_q(\frac r{r+1})+\frac 1 {r+1}\log_q(\frac 1{r+1})-\frac r{r+1}\log_q r\\
&= & 1-\frac r{r+1}\log_q(r+1)-\frac 1{r+1}\log_q(r+1)\\
&= & 1-\log_q(r+1).
\end{eqnarray*}
To attain the bound we give an algorithm that asymptotically can transmit $\tilde{M}_r(n,t)$ messages and therefore show
that the bound (\ref{generalupper_r}) can be attained. 

We start with the case $\tau>\frac r{r+1}$. 
If $\frac q{r+1}$ is an integer, we have a simple algorithm $A_r$.
We can find $\frac q{r+1}$ symbols in such a way that the receiver can distinguish them, even in the case of error. By using these symbols,
we have an algorithm with the rate 
\[
1-\log_q (r+1).
\]
For $q\geq 2(r+1)$, this algorithm can also be used if  $\frac q{r+1}$ is not an integer with a positive rate that does not reach the upper bound (\ref{generalupper_r}).
By applying Lemma~\ref{D_r}, we get an algorithm which reaches that bound.

For the case $0\leq \tau\leq \frac r{r+1}$, we can use algorithm $L_r(\Delta)$ of Lemma~\ref{L_r} because the assumptions
are fulfilled by the previous considerations. We get the rate $1- h(\tau )\log_q (r+1)$.\qed

%All main ideas of the proofs were given in the sketch. In a more detailed analysis we checked that the asymptotic behavior is correct.
%We checked that we did not loose to much when we took the logarithm, because the $k_i$ have to be natural numbers. Therefore we might to round
%p. We showed that this is not relevant for $n$ to infinity.
We use Lemma~\ref{D_r} and Lemma~\ref{L_r} to give the capacity error function
for $\Lambda_r(q)$ in Theorem~\ref{two}. Now we consider the larger set $\Gamma_r(q)$. The key point in Theorem~\ref{two} is the fact that we can 
transmit with a positive rate for any $\tau$, 
if in the graph we have at least two points at the sender side that do not overlap at the receiver side. We say that we can 
separate these two symbols. In this case we can use
the two symbols to transmit the data. 
The idea now is to find a condition for the channels in $\Gamma_r(q)$ such that we can separate two points.
We use a well-known fact to show this. If we consider the matrix representation of the 
channel, we can separate two points, if for $0\leq j_1,j_2\leq q-1$ there are two rows $a^{j_1}=(a^{j_1}_0,\dots a^{j_1}_{q-1})$ and $a^{j_2}=(a^{j_2}_0,\dots a^{j_2}_{q-1})$
such that there is no $i$ with $a^{j_1}_i=a^{j_2}_i$ for $0\leq i\leq q-1$. This known method (see \cite{C14}) to calculate this number is the key idea of the following proposition.
\begin{Proposition}\label{general}
Let $q>r^2+r+1$, then for any $0\leq \tau\leq 1$ there is an algorithm such that the rate is positive.
\end{Proposition}
{\bf Proof:} In total we have $q{r+1\choose 2}$ pairs of $(1,1)$, which means two columns with a 1 at the same position. Iff we have such pairs for
any two columns then this number is $\geq {q\choose 2}$. Therefore the proposition holds.\qed

With this proposition we can prove the following theorem.
\begin{Theorem}
Let $q>r^2+r+1$ and $\Delta\in \Gamma_r(q)$, then $\Delta$ has the following capacity error function.
\begin{eqnarray*}
C_{\Delta,f}^q(\tau )=
\begin{cases}
1- h(\tau )\log_q 2-\tau \log_q r & 
\text{if } 0\leq \tau \leq \frac{r}{r+1}\\
1-\log_q(r+1) & \text{if } \tau > \frac {r}{r+1}.
\end{cases}
\end{eqnarray*}
\end{Theorem}
{\bf Proof:} By Proposition~\ref{general} we know that we can separate two points. Therefore
there is an algorithm rate $\log_q 2$. Applying Lemma~\ref{L_r} and Lemma~\ref{D_r} we now have 
an algorithm with rate 
\begin{eqnarray*}
C_{\Delta,f}^q(\tau )=
\begin{cases}
1- h(\tau )\log_q 2-\tau \log_q r & 
\text{if } 0\leq \tau \leq \frac{r}{r+1}\\
1-\log_q(r+1) & \text{if } \tau > \frac {r}{r+1}.
\end{cases}
\end{eqnarray*}
\qed
\section{Conclusions}
The capacity error function is equal to the zero error capacity of Shannon \cite{S56} if we consider $\tau=1$. In this case any number of errors are possible.
Shannon represented the channel as a linear graph. The vertices of the graph are labeled by the code symbols.
There is an edge between the two vertices if there exist errors such that the receiver can receive the same symbols when the sender
sends one of the symbols. Shannon showed that the zero-error capacity is zero if and only if there is an edge between every pair of
vertices; otherwise the zero-error capacity is positive. If we have a channel of $\Gamma_r(q)$, we only have to check the corresponding 
Shannon graph. If there are two vertices without an edge, we can apply our method and get an algorithm.

In \cite{BET18}, the zero-error capacity of wraparound channels with limited
magnitude is considered, without feedback. 
In this paper, especially the channels for which we give the
error capacity function are considered without feedback. In these cases, it makes sense to distinguish between systematic and non-systematic errors.  
A systematic code is any error-correcting code in which the input data is embedded in the encoded output. In the case of feedback, we are not aware of any definition of a systematic code. Nevertheless, one could call our encoding algorithm $L_r$ systematic, because the receiver can reconstruct the message from the first $a$ symbols directly if no error occurs. However, this only applies to this part of the algorithm. The entire coding algorithm does not have this property.
Another interesting fact, comparing the cases without and with noiseless feedback, is Lemma~\ref{L_r}. In the case with feedback, we could also show that if $\frac{q}{r+1}$ is 
not an integer, we can give the error correcting capacity function. For $\tau=1$ this is the zero-error capacity. Shannon \cite{S56} could already show that the 
zero error capacity is known with noiseless feedback, and without noiseless feedback it is not known. It is still an open problem to get the zero-error capacity in this case (see \cite{GKV94}).
In \cite{BET18} it is shown that the zero-error capacity could be reached with a systematic code if $\frac{q}{r+1}$ is an integer. For $\tau>\frac r{r+1}$ where $\frac{q}{r+1}$ is an integer,
our algorithm also does not use the feedback, but is still optimal. 	 

Our algorithms have the property that the sender does not need to get the received symbol of the receiver immediately (for example in $L_q$). In this case such
an algorithm is called an algorithm using partial noiseless feedback.

Each $(n,M,t)_f$ error-correcting code with feedback is equivalent to the so-called Renyi-Ulam game of \cite{U76}.
This means each code gives an algorithm for the game and the other way round. For more details,
see for example, \cite{cicalese} or \cite{deppe}. Therefore, our algorithms
also give strategies to
win a generalized $q$-ary version of the Ulam game with special rules for the lies.

Some results of this paper were presented at the ACCT 2018 in Kaliningrad. 
%It might be interesting to consider unidirectional error correcting codes with feedback.
\section*{Acknowledgment}
Christian Deppe was supported in part by the Bundesministerium für Bildung und Forschung (BMBF) via grant no. 16KIS0856 and
Vladimir Lebedev was supported in part by the Russian Foundation for Basic Research, project no. 19-01-00364.

\end{document}